# Nano-Thermoelectric Infrared Bolometers

Aapo Varpula[a)], Kirsi Tappura, Jonna Tiira, Kestutis Grigoras, Olli-Pekka Kilpi, Kuura Sovanto, Jouni Ahopelto, and Mika Prunnila[a)]

VTT Technical Research Centre of Finland Ltd, Tietotie 3, FI-02150 Espoo, Finland
[a)] Authors to whom correspondence should be addressed: aapo.varpula@vtt.fi and mika.prunnila@vtt.fi

Abstract

Infrared (IR) radiation detectors are used in numerous applications from thermal imaging to spectroscopic gas sensing. Obtaining high speed and sensitivity, low-power operation and cost-effectiveness with a single technology remains to be a challenge in the field of IR sensors. By combining nano-thermoelectric transduction and nanomembrane photonic absorbers, we demonstrate uncooled IR bolometer technology that is material-compatible with large-scale CMOS fabrication and provides fast and high sensitivity response to long-wavelength IR (LWIR) around 10 μm. The fast operation speed stems from the low heat capacity metal layer grid absorber connecting the sub-100 nm-thick n- and p-type Si nano-thermoelectric support beams, which convert the radiation induced temperature rise into voltage. The nano-thermoelectric transducer-support approach benefits from enhanced phonon surface scattering in the beams leading to reduction in thermal conductivity, which enhances the sensitivity. We demonstrate different size nano-thermoelectric bolometric photodetector pixels with LWIR responsivities, specific detectivities and time constants in the ranges 179 – 2930 V/W, 0.15 – 3.1·$10^8$ cmHz$^{1/2}$/W and 66 – 3600 μs, respectively. We benchmark the technology against different LWIR detector solutions and show how nano-thermoelectric detector technology can reach the fundamental sensitivity limits posed by phonon and photon thermal fluctuation noise.

## 1 Introduction

Infrared (IR) part of the electromagnetic spectrum, where the photon wavelength is longer than that of visible light, is relevant for numerous applications ranging from thermal imaging for night vision to remote temperature measurements and chemical analysis using infrared spectroscopy [1–3]. IR imaging and spectroscopy can be utilized, for example, in the detection of cancerous cells [4], thermography in medicine, biology, and sports [5,6] as well as in industrial applications such as bioprocess monitoring [7], and in dynamic material studies [8]. The thermal emission of most objects in our ambient is strongest in the long-wave infrared range (LWIR), typically at wavelengths 8–15 μm, thereby making this range ideal for IR applications as fully passive operation can be achieved.

In the IR range, the most common detector types are quantum detectors based on electron-hole pair generation and thermal detectors (bolometers) based on radiation-induced heating of an absorber. Quantum IR detectors, either photovoltaic or photoconductive, typically have higher sensitivity than bolometers, but require cooling from room temperature down to cryogenic temperatures to reach the maximum performance, especially, in the LWIR range where photon energies are relatively small. The quantum detectors for the LWIR range require also relatively expensive and toxic materials, such as HgCdTe and InAsSb. As quantum detectors transduce the IR signal directly into electric, they are typically faster than the state-of-art thermal detectors. A high-speed detector not only provides fast data acquisition, but also enables imaging and sensing of fast processes and environments, which is a necessity, for example, in vehicular applications.

The state-of-the-art bolometers can reach relatively high sensitivities without cooling. The bolometers are essentially thermometers, which detect the temperature change of the absorber. Widely used resistive bolometers [1,9,10] are based on temperature-dependent resistors. Diode and transistor based bolometers [11–13] are similar to resistive bolometers, but they use semiconductor devices as sensitive thermometers. In thermoelectric bolometers [1,9,10], often also referred to as thermopiles or thermocouples, the thermoelectric transduction of thermal signals into electrical signals has, in principle, low noise, as the main noise sources are the fundamental thermal fluctuation noise and the Johnson-Nyquist noise [9,10,14]. Furthermore, the thermoelectric transduction does not require any external power in signal generation, thus enabling low power operation. Although here we focus on detectors operating at room temperature, we note that interesting thermoelectric detector concepts operating at cryogenic temperatures have also been proposed [15,16].

Traditional thermoelectrics relies on the bulk material properties, whereas, nano-thermoelectrics capitalizes on different combinations of nano-fabrication and synthesis techniques to engineer the electro-thermal performance, i.e., charge carrier and phonon transport [17–20]. Nano-thermoelectrics has been mainly driven by applications in thermal energy harvesting [21–24] with very little attention to detectors until recently, when it was postulated that nano-thermoelectrics can provide an attractive low noise transduction method for bolometers [14,25].

Silicon is an attractive detector material as it is widely used in semiconductor industry, cost-efficient, non-toxic, and, in general, abundantly available. Shaping silicon to nano-scale membranes [26,27] or wires [28] causes its thermal conductance to collapse due to increased phonon scattering while keeping the Seebeck coefficient and electrical conductivity virtually unaltered. This increases the thermoelectric figure of merit significantly and, together with the maturity of the silicon technology, make Si-based nanostructures extremely attractive transducer materials for thermoelectric detectors.

For efficient radiation-to-electricity conversion, bolometers require an efficient radiation absorber. To enhance the performance, it is essential that the absorber is supported solely by the thermoelectric legs to reduce the thermal coupling to the surroundings. Various optical absorbers based on thin layers have been proposed and demonstrated for bolometers [29]. The thermal mass, or heat capacity, of the absorber together with the thermal conductance to the surroundings defines the speed of the bolometer. Therefore, for example for imaging applications, high absorption efficiency should be obtained with minimal absorber heat capacity.

In this work, we demonstrate for the first time nano-thermoelectric infrared detector pixel technology. In nano-thermoelectric photodetectors a high efficiency absorber with low heat capacity is supported solely by the thermoelectric transducer beams (see Fig. 1). The thermoelectric beams are heavily doped poly-Si nanolayers. The thin-film metal grid acts, in addition to being the radiation absorber, as the electric contact between the n- and p-type beams forming the thermocouples. We demonstrate the use of both TiW and TiN as the metal-grid based effective media. These material choices make the detector cost-effective by allowing standard fabrication processes. The technology presented here is easily scalable to various sizes of matrices and single pixel detectors. For our TiW and TiN based devices with ~25 x 25 µm² absorbers we obtain very small thermal time constants $\tau$ of 66 and 190 µs and IR responsivities of 179 and 494 V/W resulting in specific detectivities of $1.5 \cdot 10^7$ cmHz$^{1/2}$/W and $8.7 \cdot 10^7$ cmHz$^{1/2}$/W. For our larger device with a 40 x 40 µm² absorber, we obtain larger sensitivity, as demonstrated by the respective figures of 2930 V/W and $3.1 \cdot 10^8$ cmHz$^{1/2}$/W, at a speed corresponding to $\tau$ = 3.6 ms. These results pave the way for scalable nano-thermoelectrics based IR detection technology for different applications from chemical sensing to thermal imaging.

## 2 Operation principles of thermoelectric bolometers

Thermoelectric bolometers can be described by a model based on lumped thermal RC circuit [10,14,25,30]. The speed of the bolometer is characterized by the thermal time constant $\tau = C_{\text{th}}/G_{\text{th}}$, where $C_{\text{th}}$ is the heat capacity of the absorber and $G_{\text{th}}$ the thermal conductance from the absorber to the environment, which is also one of the determining factors of the bolometer sensitivity. $G_{\text{th}}$ is a sum of the conductance stemming from two different heat transfer mechanisms, conduction and radiation, where in the case of uniform absorber, the latter can be expressed as $G_R = 4A_{\text{abs}}\epsilon\sigma T^3$, where $A_{\text{abs}}$ is the area of the absorber, $\epsilon$ the emissivity, $\sigma$ the Stefan-Boltzmann constant, and T absolute temperature. The thermoelectric transducer of the bolometer transforms the temperature gradient between the absorber and the bath into voltage, and is characterized by the total Seebeck coefficient $S = dV/dT = S_{\text{p}} - S_{\text{n}}$, where $S_{\text{p}}$ and $S_{\text{n}}$ are the Seebeck coefficients of the p- and n-type thermoelectric elements.

The optical performance of the bolometer is determined by the wavelength-dependent optical (spectral) efficiency $\eta(\lambda)$ of the absorber. As the optical efficiencies of absorbers generally depend on the wavelength, the total optical efficiency $\eta_{\text{tot}}$ of a bolometer is specific to the spectrum of the optical power. This optical efficiency can be written in terms of the absorbed optical power P$_{\text{abs}}$ and the total optical power P incident on the absorber as $\eta_{\text{tot}} = \frac{P_{\text{abs}}}{P} = \frac{\int \eta(\lambda) P_\lambda(\lambda) d\lambda}{\int P_\lambda(\lambda) d\lambda}$, where $P_\lambda(\lambda)$ is the spectral incident optical power at wavelength $\lambda$. The optical modelling of the absorber is described below.

The frequency dependence of the output voltage amplitude of a thermoelectric bolometer is given by (cf. [14,25,30])

$$V = \frac{V_{\text{ampl}}}{\sqrt{1 + \tau^2 \omega^2}}, \qquad (1)$$

where $\omega$ is the angular frequency of the optical power and $V_{\text{ampl}} = S\eta_{\text{tot}}P/G_{\text{th}}$. The phase of the output voltage is

$$\theta = \arctan(-\tau\omega). \qquad (2)$$

Below the thermal cut-off angular frequency of $\omega_c = 1/\tau$, eq. (1) gives $dV/dP = S\eta_{\text{tot}}/(G_{\text{th}}\sqrt{1+\tau^2\omega^2})$, and furthermore the (voltage) responsivity of the bolometer as

$$R_V = \left.\frac{dV}{dP}\right|_{\omega \ll \omega_c} = \frac{S\eta_{\text{tot}}}{G_{\text{th}}}. \tag{3}$$

The sensitivity of a bolometer (or any photodetector) is often described by specific detectivity D* which relates to the optical noise-equivalent power (NEP) by $D^* = \sqrt{A_{\text{abs}}}/\text{NEP}$, where NEP is defined for a unit bandwidth of 1 Hz. In thermoelectric bolometers, the main intrinsic noise sources are the Johnson-Nyquist noise associated with the total electrical resistance R of the thermoelectric transducer of the bolometer, and the thermal fluctuation noise. The latter originates from random energy exchange between the absorber of the bolometer and its surroundings by different heat carriers (electrons, phonons and photons). The corresponding optical NEP of the thermal fluctuation noise is [9]

$$\text{NEP}_{\text{th}} = \frac{\sqrt{4k_B T^2 G_{\text{th}}}}{\eta_{\text{tot}}}. \tag{4}$$

where $k_B$ is Boltzmann's constant. For the Johnson-Nyquist noise this is given by the ratio of the Johnson-Nyquist voltage noise $v_{\text{JN}} = \sqrt{4k_B TR}$ and the frequency-dependent bolometer responsivity as $\text{NEP}_{\text{JN}}(\omega) = (dP/dV)\sqrt{4k_B TR} = R_V^{-1}\sqrt{4k_B TR}\sqrt{1+\tau^2\omega^2}$. Below the thermal cut-off ($\omega \ll \omega_c$) this reduces to

$$\text{NEP}_{\text{JN}} = \frac{\sqrt{4k_B TR}}{R_V}. \tag{5}$$

These two independent noise sources can be combined into total optical NEP as (cf. [14,25,30])

$$\text{NEP} = \sqrt{\text{NEP}_{\text{th}}^2 + \text{NEP}_{\text{JN}}^2} = \text{NEP}_{\text{th}}\sqrt{1+\frac{1}{\widetilde{ZT}}}, \tag{6}$$

where $\widetilde{ZT} = S^2 T/(G_{\text{th}} R)$ is the effective thermoelectric figure of merit of the bolometer. $\widetilde{ZT}$ coincides with the material thermoelectric figure of merit ZT when the geometries of the n- and p-thermoelectric elements and the absolute values of the material parameters are equal [14]. Here eq. (6) is defined for the frequencies below the thermal cut off. The frequency dependency of eq. (6) can be restored by performing substitution $1/\widetilde{ZT} \rightarrow (1+\tau^2\omega^2)/\widetilde{ZT}$. Eq. (6) shows that the fundamental thermal fluctuation noise dominates when $\widetilde{ZT} > 1/3$, and the Johnson-Nyquist noise when $\widetilde{ZT} < 1/3$. NEP can be minimized by maximizing $\widetilde{ZT}$ and $\eta_{\text{tot}}$ and by minimizing $G_{\text{th}}$. The maximal $\widetilde{ZT}$ can be achieved by optimizing the geometries of the thermoelectric elements as well as their material parameters, i.e., Seebeck coefficients and electric and thermal conductivities [14].

## 2.1   Optical modelling of absorber

The absorption of infrared radiation in the present nanobolometers is based on an electrically conducting absorber film (either TiW or TiN), which is patterned into a grid to form an effective medium. This absorber grid is coupled to an optical cavity with a back reflector at the bottom of the cavity (see description in Section 3).

The optical properties of the absorber structure are calculated with full-wave electromagnetic simulations. The three-dimensional Maxwell equations are solved using the finite element method (FEM) software Comsol Multiphysics® [31] with perfectly matched layer (PML) boundary conditions. Periodic boundary conditions perpendicular to the plane of the absorber membrane are used to generate a rectangular periodic lattice of the defined unit cell geometry. The incident radiation is simulated by plane waves that propagate normal to the surface of the detector. The dispersive permittivity of the absorber metal is derived from the Drude model [32,33] with the values of the electrical conductivity of $9.2 \times 10^5$ S/m and $7.57 \times 10^5$ S/m measured for the thin TiW and TiN films, respectively, and assuming the carrier relaxation time to be $1.5 \times 10^{-15}$ s [34]. Similarly, the permittivity of polysilicon was obtained based on the Drude model, but here the doping concentration and mobility were used as the input parameters [35]. The doping concentration of the silicon substrate was $1 \times 10^{16}$ cm$^{-3}$. The time-average spectral power density, $Q_{abs}$, absorbed in volume V is used to calculate the power absorbed in the absorber membrane [36]:

$$Q_{abs}(\omega) = \frac{\omega}{2} \int_V Im[\varepsilon(\omega)]|\boldsymbol{E}|^2 dV, \qquad (7)$$

where E is the electric field vector and $\varepsilon(\omega)$ the dispersive permittivity, given by the Drude model, of the material in volume V.

## 3  Novel IR bolometer structure

Scanning electron microscope images of the fabricated nano-thermoelectric bolometers are shown in Fig. 1 together with diagonal schematic cross-sections and a summary of device characteristics. We demonstrate the realization of the same concept using two sets of materials. The devices consist of an optical absorber supported by n- and p-doped 80 nm (Dev. A) or 70 nm (Devs. B and C) thick poly-Si beams forming two thermocouples and a frame controlling the strain in the beams [37]. In Device A, this frame is silicon nitride located below the poly-Si while in Devices B and C, the frame is atomic layer deposited Al$_2$O$_3$ on the poly-Si. The thermocouple pairs are electrically connected by a 30 nm (Dev. A) or 50 nm (Devs. B and C) thick layer of metal, which at the same time acts as the absorbing element. In Device A, the absorber metal is TiW, and in Devices B and C it is TiN. This metal-poly-Si stack absorber is patterned into a grid to control the optical impedance of the absorber, and for easier release of the suspended parts of the device during fabrication. Under the absorber grid an optical cavity is formed between the absorber and the substrate which acts as the back reflector. The optical cavity depth of 2.5 µm was selected to maximize the detector output for room-temperature thermal radiation with the wavelength maximum around 10 µm. Optically the absorber, the optical cavity, and the back reflector form a well-known quarter-wave resistive absorber, also known in radio-frequency engineering as the Salisbury screen, where the sheet resistance of the absorber is matched to the vacuum impedance [38]. This kind of structure is also known as an antiresonant interference structure [29].

## 4  Experimental details

### 4.1  Device fabrication

The devices were fabricated in the VTT Micronova cleanroom facilities on a 150 mm standard single-side-polished p-type (1-50 Ωcm) silicon wafer. First, a 2.5 µm thick layer of sacrificial SiO$_2$ was deposited by a tetraethyl orthosilicate (TEOS) low pressure chemical vapor deposition (LPCVD) process. Next, in the case of Device A, a 270-nm-thick stress-compensation SiN$_x$ layer was deposited by LPCVD. This SiN$_x$ layer was patterned by plasma etching to form a strain-compensation frame similarly to in our previous work [14,25,37]. Then, in all devices, 80 nm (Dev. A) or 70 nm (Devs. B and C) of LPCVD polysilicon, was deposited. The poly-Si was doped selectively with boron and phosphorus by ion implantation and annealing at 700 °C for 2 hours. Resistivities of 5.3 mΩcm and 3.4 mΩcm, charge-carrier mobilities of 27 cm$^2$/Vs and 15 cm$^2$/Vs, and Hall carrier concentrations of $4.4 \cdot 10^{19}$ cm$^{-3}$ and $1.3 \cdot 10^{20}$ cm$^{-3}$ were obtained for the n- and p-type 80 nm thick poly-Si, respectively, using van der Pauw structures. For the 70 nm thick poly-Si, the corresponding values are 4.7 mΩcm and 3.0 mΩcm, 24 cm$^2$/Vs and 14 cm$^2$/Vs, and $5.5 \cdot 10^{19}$ cm$^{-3}$ and $1.4 \cdot 10^{20}$ cm$^{-3}$. These results are in line with the data in the literature for poly-Si films [39,40]. The poly-Si layer was patterned by plasma etching to form the thermoelectric beams. After the poly-Si processing, in the case of Devices B and C, a 40 nm of $Al_2O_3$ was deposited by atomic layer deposition (ALD) technique. After that, all the devices followed the same process flow. The absorber and contact metal, 30 nm of TiW (Dev. A) or 50 nm of TiN (Devs. B and C), and 300 nm Al pad metal, were sputtered and patterned by plasma and wet etching steps. Finally, the devices were released with HF vapor etching of the sacrificial oxide layer.

### 4.2 Infrared characterization

The optical characterization of the fabricated detector was performed in a vacuum chamber with pressure < 1 Pa by illuminating the sample using a calibrated cavity blackbody infrared source (model SR-200 33 of CI-Systems, emissivity 0.99 ± 0.01, temperature accuracy ±2 K) through a ZnSe window of the vacuum chamber lid. The detector was kept at room temperature with a temperature controlled sample holder. The total optical power is controlled by the area of the variable circular output aperture and the temperature of the blackbody IR source. The temperature of the blackbody IR source determines the spectrum of the IR radiation incident on the detector (see Suppl. Fig. 2 in Supplementary Material). In the setup, the spectrum obtained with the blackbody temperature of 200 °C suits rather well for the present detectors as majority of the IR spectrum is within the optimal absorption range of the detectors. Higher blackbody temperatures provide stronger IR signal (and thereby better accuracy) at all wavelengths, but shift the radiation peak to lower wavelengths.

To ensure that the measured signal is not caused by photovoltaic effects in silicon, a high-resistivity (>5 kΩcm) double-side-polished silicon wafer was used as an optical filter directly in front of the ZnSe window. As Si is transparent above the wavelength of ~1.1 µm, the Si filter has only a minor effect on the shape of the mid and long-wave infrared spectrum used in the experiments. The combination of the ZnSe window and the high resistivity Si filter results in a relatively flat transmission in the wavelength range of 1–20 µm (see Supplementary Material). The total IR power incident on the detector absorber through the optical system is estimated using a numerically calculated blackbody spectrum and the analytical model of the optical setup with a Lambertian blackbody source (cf. [41]) as described in the Supplementary Material.

The opto-thermoelectric response of the detector was measured with a lock-in technique using either an optical chopper or shutter, a SR865A lock-in amplifier and a SR560 preamplifier connected to the detector in a differential mode. In the measurement, the two thermocouples of the bolometer were connected in parallel, i.e., the n+ ends of the two thermocouples were connected together as well as the p+ ends (see Fig. 1). The optical shutter was used at very low frequencies (5 Hz and below). In high-frequency response measurements (1 kHz and above), an optical chopper wheel with smaller slits, and thus, a small output aperture for the infrared radiation is needed. Therefore, the optical power was maximized by using a high blackbody temperature.

# 5 High-speed and sensitive detectors

The measured frequency responses of the IR detectors of Fig. 1 are shown in Fig. 2. The model of eqs. (1) and (2) fits well both to the experimental amplitude and phase data. We can observe that Device A is extremely fast: The model fit yields the detector thermal time constant of 66 µs, which is exceptionally small for a thermal detector. The corresponding thermal cut-off frequency $\omega_c/2\pi$ is 2.4 kHz suggesting that the detector is capable of detecting optical signals with frequencies up to several kHz depending on the power of the measured signal. Device B is also fast with the thermal time constant of 190 µs (see Table 1 below). As the device performance is always a compromise between the speed and responsivity, the responsivities of Device B and C are much larger than those of Device A. For example, Device C is slowest and has the highest responsivity due to its longest beams. The largest absorber size of Device C reduces its speed as well.

Figure 3a shows the response of the detectors to blackbody infrared signals with power levels extending over two orders of magnitude. The responsivity is measured at low frequencies well below the thermal cutoff. The response data shows good linearity over the whole power range, thereby demonstrating excellent dynamic range of the detectors. The characteristics of the present bolometers are summarized in Table 1. Compared to Device A, the higher sensitivities of Devices B and C are clearly shown in Fig. 3a. The spectral efficiencies of the absorbers $\eta(\lambda)$, simulated using circularly polarized IR radiation, are shown in Fig. 3b. The response of these kinds of absorbers is independent of the polarization (see Supplementary Material). As all the detectors in Fig. 3b exhibit relatively similar spectral efficiencies, the larger responsivities of Devices B and C can be explained by their thinner poly-Si, the lower thermal conductivity of which generates a higher thermal signal. In Fig. 3b the spectral efficiencies of the absorbers $\eta(\lambda)$ peak around the wavelength of 10 µm, thus making the present bolometers ideal for thermal signals from bodies around room temperature. The peak of the blackbody radiation shifts to lower wavelengths at higher blackbody temperatures (e.g. 6.1 µm at 200 °C and 2.7 µm at 800 °C), which results in a slight reduction of the responsivity of the detectors. However, the intensity of the blackbody radiation increases at all wavelengths and, thereby, the total detected signal grows as a function of the temperature of the blackbody regardless of the optimal wavelength range of the detector. Thus, the blackbody temperatures selected for the measurements give a good basis for the performance characterization covering a broad range of wavelengths (see Supplementary Figure 2) and providing a good accuracy with a relatively simple measurement setup.

The finite element method (FEM) simulations of the poly-Si-metal absorber grid showed that both the TiW and TiN layers govern the optical absorption process in the bolometers, and the poly-Si layer below the absorber metal layers is mostly inactive, because its electric conductivity is too low at IR frequencies. The optical impedance matching of the absorber is determined by the sheet resistance of the absorber metal layer. In all the devices, the effective surface impedances of the absorber are below of the optimal 377 Ω (~ 287 Ω in Device A, and ~ 132 Ω in Devices B and C). The optimal impedance matching can be achieved by tuning the geometry of the absorber grid or the thickness of the absorber metal layer [42]. Another way to improve the absorptance is to increase the reflectance of the back reflector from the Si substrate value of about 30%, for example by introducing a metal layer on top of the substrate in the cavity in the beginning of the fabrication process, or using a heavily doped substrate in the fabrication. These straightforward improvements would allow the absorptivity to be increased virtually to 100% for a selected wavelength range. For example, in the case of the TiN absorber (employed by Devices B and C), high absorption efficiency can be obtained by thinning the TiN layer to 24 nm and introducing a perfect mirror as a back reflector, as shown in Fig. 3b.

Eq. (3) allows the estimation of the material parameters of the thermoelectric transducers of the detectors. The literature data on LPCVD polycrystalline silicon with similar resistivities as in the present work [39,43] suggests that the Seebeck coefficients of the n- and p-type poly-Si are –0.25…–0.30 mV/K and 0.2 – 0.25 mV/K, respectively. Based on these values, the total Seebeck coefficient S of the detectors is 0.45 – 0.55 mV/K, which is consistent with our previous results [25]. For the estimations of the 80 nm poly-Si we can use the data of Device A. Using the 200 °C value of $R_V$ = 179 V/W (see Table 1) and the simulated total optical efficiency of absorber in the characterization conditions, $\eta_{tot}$ = 0.48, eq. (3) gives $S/G_{th}$ = 373 V/W. Furthermore, with the estimated S, we obtain 1.2 – 1.5 µW/K for the thermal conductance $G_{th}$ of Device A. For Devices B and C with 70 nm poly-Si and $\eta_{tot}$ of 0.46, we obtain $G_{th}$ values of 0.41 – 0.51 µW/K, and 0.07 – 0.09 µW/K, respectively. The thermal conductances $G_{th}$ of Devices B and C are much smaller than that of Device A due to the lower thermal conductivity of the thinner poly-Si. In the case of Device C, $G_{th}$ is further reduced due to the longer thermoelectric beams. The thin-film thermal conductivities of the absorber metals (in the metal/poly-Si absorber stack) range typically from 60 W/(mK) of TiW [44] to 1.2 – 4.7 W/(mK) of TiN [45,46]. Here, it should be noted that although there is a thermal boundary between the poly-Si and the absorber metal, the typical Si-metal thermal boundary resistances [45,47,48] are insignificantly small.

By assuming that thermal conductivities of the n- and p-type poly-Si beams are equal in Devs. A and B, we estimate that the thermal conductivities of 70 nm and 80 nm poly-Si are 7 – 8 W/(mK) and 17 – 21 W/(mK), respectively. These values are around an order of magnitude smaller than in bulk silicon. A low thermal conductivity is a prerequisite for turning silicon into a material with a high thermoelectric figure of merit. The thermal conductivity estimates are in line with the thermal conductivities of undoped and doped single and polycrystalline silicon membranes with similar thickness ranges [14,25–27,40]. Thicker poly-Si films with higher doping levels than in the present poly-Si membranes and bulk poly-Si exhibit thermal conductivities in the range of 30 – 120 W/(mK) [43,49–51]. The thermal conductivity of the present poly-Si membranes suggests that phonon surface scattering reduces the thermal conductivity similarly as in single-crystalline Si nanomembranes [26,27].

The electrical resistance R of the thermoelectric bolometer is crucial to the sensitivity of the device as it determines the intrinsic voltage noise of the detector [14]. The resistance of a real device includes additional series resistance that is not directly originated from the thermoelectric transducer. To clarify this, we describe the total measured resistance of the bolometer as $R_{tot}$. In the present bolometers, the resistance of the absorber grid has ideally only little contribution to the total resistance of the devices as the effective sheet resistance of the absorber grid is close to the vacuum impedance (~377 Ω). In the present type of devices, the additional series resistance originates, for example, from the resistance of the polysilicon–metal contact in the absorber grid. In Device A, this poly-Si-TiW contact resistance dominates: the total resistance of the thermoelectric transducer is $R_{tot}$ = 50 kΩ, and only a small portion of it, namely $R_{beams}$ = 2.5 kΩ, arises from the thermoelectric poly-Si beams. In Device B the effect of the contact resistance is much smaller, and in Device C, the beam resistance $R_{beams}$ is dominating $R_{tot}$. The effect of the R values on the device performance can be clearly seen in $\widetilde{ZT}$. For Device A, the values of R listed in Table 1 and the value of $G_{th}$ estimated above give $\widetilde{ZT}$ = 1.0–1.2·10$^{-3}$ using the measured total bolometer resistance $R_{tot}$, and $\widetilde{ZT}$ = 0.020–0.025 using the resistance of the thermoelectric beams, $R_{beams}$, only. For Devices B and C, the corresponding former values are an order of magnitude larger, $\widetilde{ZT}$ = 0.010–0.015, and the latter $\widetilde{ZT}$ = 0.015–0.067.

Table 1. Summary of the characteristics of the nano-thermoelectric IR bolometers of Fig. 1.

| Device | A | B | C |
|---|---|---|---|
| Thickness of thermoelectric poly-Si | 80 nm | 70 nm | |
| Absorbing material | TiW 30 nm | TiN 50 nm | |
| Absorber fill factor $\eta_{FF} = A_{abs}/A$ | 66.4% | 67.3% | 5.43% |
| Measured thermal time constant $\tau$ [µs] | 66 ± 3 | 190 ± 1 | 3596 ± 7 |
| Total bolometer resistance $R_{tot}$ [kΩ] | 50.0 | 11.9 | 86.4 |
| Resistance of thermoelectric beams $R_{beams}$ [kΩ] | 2.5 | 2.7 | 58.1 |
| Measured responsivity $R_V$ (200 °C BB radiation) [V/W] | 179 ± 10 | 494 ± 28 | 2930 ± 93 |
| Measured responsivity $R_V$ (800 °C BB radiation) [V/W] | 157 ± 4 | 354 ± 9 | 2276 ± 64 |
| Optical noise equivalent power[a,b] (NEP) [pW/Hz$^{1/2}$] | 160 ± 9 | 28 ± 2 | 13 ± 1 |
| Optical NEP determined by $R_{beams}$ only[a,c] [pW/Hz$^{1/2}$] | 36 | 13 | 11 |
| Estimated optical NEP with $R_{tot}$ and 100% absorption[a,b,d] [pW/Hz$^{1/2}$] | 76 | 13 | 5.8 |
| Estimated optical NEP with $R_{beams}$ and 100% absorption[a,c] [pW/Hz$^{1/2}$] | 17 | 6.1 | 4.8 |
| Specific detectivity[a,b] D* [10$^6$ cmHz$^{1/2}$/W] | 15 ± 0.8 | 87 ± 5 | 309 ± 10 |
| D* determined by $R_{beams}$ only[a,c] [10$^8$ cmHz$^{1/2}$/W] | 0.69 | 1.84 | 3.77 |
| Estimated D* with R and 100% absorption[a,b] [10$^8$ cmHz$^{1/2}$/W] | 0.32 | 1.92 | 6.79 |
| Estimated D* with $R_{beams}$ and 100% absorption[a,c] [10$^8$ cmHz$^{1/2}$/W] | 1.43 | 4.03 | 8.28 |

[a]Sensitivity to 200 °C blackbody (BB) radiation.
[b]Sensitivity determined by total resistance, which is limited by additional series resistance such as contact resistance. Calculated using eq. (5) with R = $R_{tot}$.
[c]Limiting sensitivity determined by the resistance of the thermoelectric beams. Calculated using eq. (5) with R = $R_{beams}$.
[d]Coincides with the estimated electrical NEP of the detector

Because the estimated $\widetilde{ZT}$ is well below 1/3, the intrinsic voltage noise of the present bolometers is determined by the Johnson-Nyquist noise, see eq. (6), Ref. [14], and experimental verification in Supplementary Material. This allows us to calculate the optical NEPs and specific detectivities D* of the detectors using eq. (5) by applying the total bolometer resistance $R_{tot}$. In the ideal case, the thermoelectric beams alone ($R_{beams}$) determine the total resistance $R_{tot}$ of the bolometer, which would decrease the NEP and increase D* from the measured values. As discussed above, the total optical efficiency $\eta_{tot}$ of the bolometer can be increased to 100% by improving the absorber efficiency. These calculated and estimated parameters are listed in Table 1. The resulting values are competitive with the state-of-the-art IR detectors (see detailed comparison below). Furthermore, $\widetilde{ZT}$ and thereby NEP and D* of the present bolometers can be improved by selecting optimal geometries for the thermoelectric beams based on the resistivities, Seebeck coefficients and thermal conductivities of the n- and p-type thermoelectric materials (see Ref. [14] and further the discussion on $\widetilde{ZT}$ below). The performance can be improved even further by geometrical optimization of $\widetilde{ZT}$.

## 6 Routes to optimize the performance

Since the surface scattering of phonons increases with decreasing nanomembrane thickness, leading to decrease in thermal conductivity [27], the performance of the thermoelectric transducer of the detector can be improved by thinning the poly-Si membrane further from the demonstrated thicknesses of 70 nm and 80 nm. Even further performance enhancement can be achieved by maximizing the power factor $S^2/R$ by optimizing the doping parameters. However, since S decreases with increasing electric conductivity [20,52], the optimization is more complicated than just minimizing R by increasing the doping. This dependence of the power factor and ZT on the doping density of single-crystalline silicon has been rather recently studied in experimental nanoscale thermoelectric generators [53].

The effect of a thinner poly-Si membrane can be estimated using the thermal conductivity of 8 W/(mK) measured for a 9 nm thick single-crystalline silicon layer with native oxide [27]. By optimizing the absorber we can reach unity $\eta(\lambda)$, as discussed above. If the grid geometries were fixed, the optimal thicknesses of the absorber metals would be 19 nm for Dev. A (with TiW) and 18 nm for Devs. B and C (with TiN) according to the electromagnetic simulations. This and the use of thinner poly-Si leads to lower heat capacity and higher speed for the devices (depending on the geometry). Furthermore, we see that the detectors can be optimized so that their resistance is determined by the resistance of the thermoelectric beams $R_{beams}$ only. This, and the optimization of the thermoelectric material would double power factor $S^2/R$, leading to $\widetilde{ZT}$ = 0.11 for Devices A-C, which leads to D* values well above $10^9$ cmHz$^{1/2}$/W for all the devices. The details of these estimations are described in Supplementary Material.

The specific detectivity vs. time constant performance of our nano-thermoelectric bolometers is plotted in Fig. 4 together with LWIR detector results reported in the literature. Note that in order to give a comprehensive view of the whole LWIR detector field in addition to thermal detectors we have also included cooled and uncooled quantum detectors in Fig. 4. Device A is an order of magnitude faster than the fastest state-of-the-art thermoelectric bolometer and 3 times faster than the fastest state-of-the art resistive bolometers. The sensitivity of the present device is in the same order of magnitude as some thermal detectors and most uncooled quantum detectors. The fill factors of Devices A and B, $\eta_{FF}$ of 66 – 67%, are large compared to the fill factors of many existing thermoelectric bolometers [54,55]. As an example, the detector in Ref. [56] has a fill factor of 1%. Detectors with a large fill factor are attractive in imaging applications where tightly packed detector arrays are needed. As discussed above, we expect that the performance of our present device can be improved remarkably by straightforward material and geometrical optimization. To highlight this, Fig. 4 also shows the performance values of the resistance-optimized nano-thermoelectric bolometers with the total resistance R determined by the beams only and unity absorptance (see Table 1) as well as the proposed optimized detectors based on 9 nm poly-Si designed either for speed or sensitivity, as described in Supplementary Material. These estimated cases cover thermal time constants over three orders of magnitude. This demonstrates the versatility of this technology and its feasibility to many applications. Note that as a rough rule of thumb for all bolometers the geography of detectivity vs time-constant plot (Fig. 4) can be interpreted so that the points with smaller (larger) time-constant and detectivity are for imaging array (single/few pixel sensing) applications.

The detector sensitivity can be enhanced further by utilizing high ZT materials [57–59], which allow NEP to be pushed to the thermal fluctuation limit defined by Eq. (4). It should be noted that the dashed horizontal line in Fig. 4 denoting the 300 K background radiation limit is also a photon thermal fluctuation noise limit, which follows from Eq. (4) in the case where only photons exchange heat between the absorber and the environment. The specific detectivity values of the largest pixels (Fig. 4, estimated points) approach this limit suggesting that nano-thermoelectric IR detectors could operate at the thermal background limit. The time constants for these detectors are relatively large excluding high speed imaging applications but allowing for single or few pixel chemical sensing applications, for example.

# 7 Conclusions

In conclusion, by combining nano-thermoelectrics and nanomembrane photonics, we have demonstrated LWIR bolometers based on CMOS-compatible materials and shown that this technology can be utilized in wide range of applications by tailoring the speed and sensitivity according to the specific needs. Nano-thermoelectric IR detectors can provide high speed and high sensitivity, low-power operation and cost-effectiveness with-in single technology. For the fabricated devices of different sizes, the measured responsivities are in the range of 180 – 2900 V/W and specific detectivities $1.5 \cdot 10^7$ – $3.1 \cdot 10^8$ cmHz$^{1/2}$/W. The thermal time constant of the fastest device is as low as 66 μs and even the device with a larger absorber and highest specific detectivity $3.1 \cdot 10^8$ cmHz$^{1/2}$/W exhibits a time constant (3.6 ms) comparable with the existing state-of-the-art bolometers. The fast operation speed stems from the low heat capacity multi-functional absorber, where a single metal layer grid absorbs the radiation and, at the same time, connects the poly-silicon n- and p-type nano-thermoelectric support beams, which convert the IR radiation induced temperature rise into voltage. In addition, we have discussed how the specific detectivity of the nano-thermoelectric bolometers can be pushed above $1 \cdot 10^9$ cmHz$^{1/2}$/W without compromising the speed of the detector and benchmarked the performance with the state-of-the art LWIR detectors.

# Supplementary material

See supplementary material for supporting details on the IR characterization, the optical modelling of absorbers, the detector noise and performance estimation, and the LWIR detectors of the comparison.

# Acknowledgements

This work has been financially supported by European Union Future and Emerging Technologies (FET) Open under Horizon 2020 programme (Grant Agreement No. 766853, project EFINED), by Business Finland co-innovation project RaPtor (No. 6030/31/2018), and by the Academy of Finland projects EXACT, BOLOSE, and LAMARS (Grant Nos. 295329, 314447, and 314809, respectively). The work of Jonna Tiira was personally supported by Academy of Finland through Grant No. 324838. The work is part of the Academy of Finland Flagship Programme, Photonics Research and Innovation (PREIN), decision 320168. We acknowledge gratefully the technical assistance of Teija Häkkinen, Kai Viherkanto, Paula Holmlund, and Tahvo Havia in device fabrication and characterization.

## Data availability

The data that supports the findings of this study are available within the article and its supplementary material.

# Figures

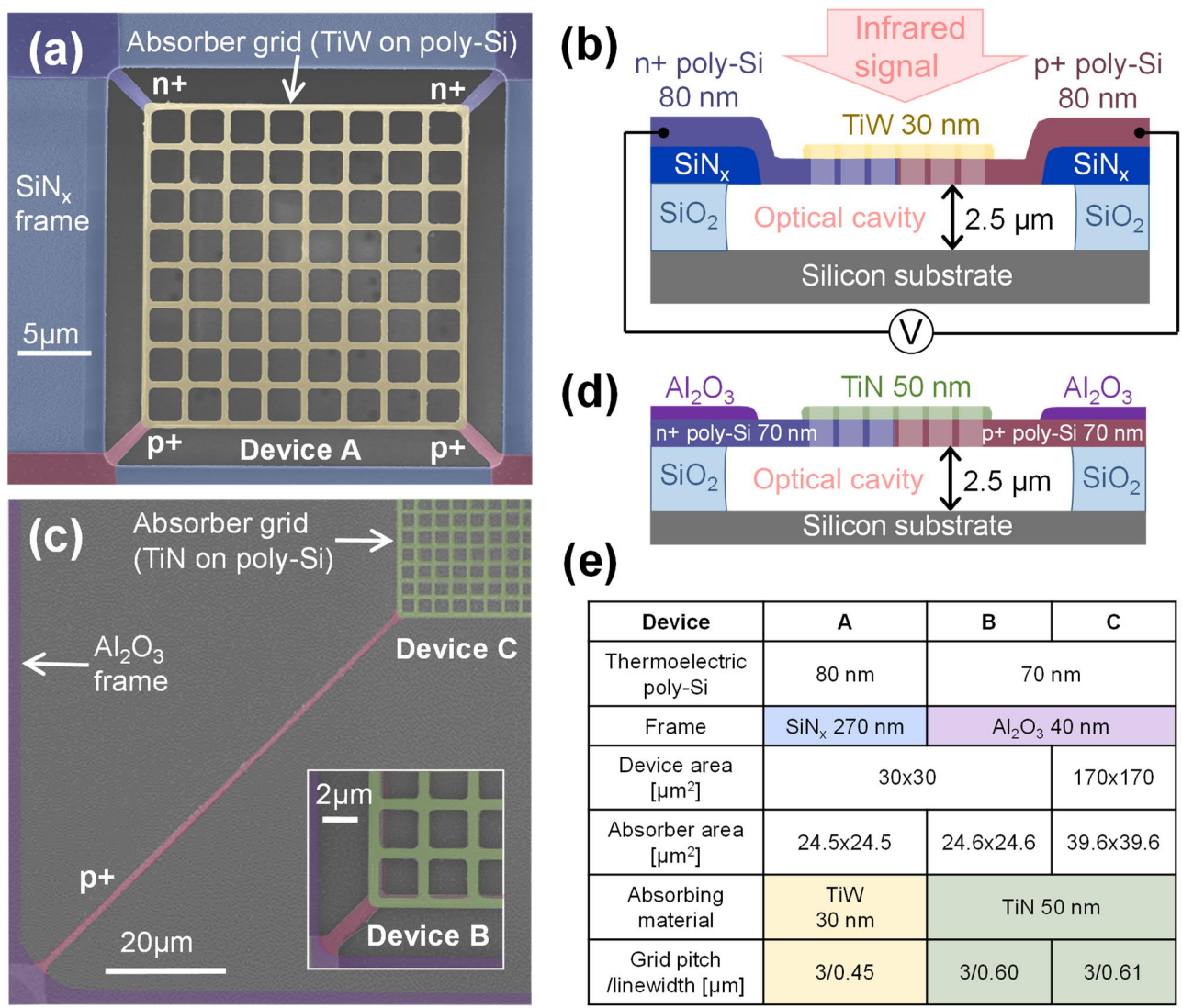

Figure 1. (a, c) Colored scanning electron micrographs of the nano-thermoelectric bolometers (top view), and (b,d) schematic diagonal cross-sections of Device A (b) and Devices B-C (d) (not in scale). The dimensional and material characteristics of the devices are listed in Table (e). The absorber grid in the middle of the devices consists of a stack of n- and p-doped, either 70 nm or 80 nm thick, polycrystalline silicon and either 30 nm thick TiW or 50 nm thick TiN absorber metal on top. The poly-Si layer extends outside the absorber grid where it is patterned into n- and p-doped beams connected to either a 270 nm thick silicon nitride or 40 nm thick $Al_2O_3$ frame suspended over the 2.5 μm deep optical cavity. The device consists of two n- and p-type poly-Si thermocouples connected in parallel by the TiW or TiN absorber metal. The absorbing materials (TiW and TiN), grid pitch, and grid linewidths of the absorbers of the devices are given in Table (e).

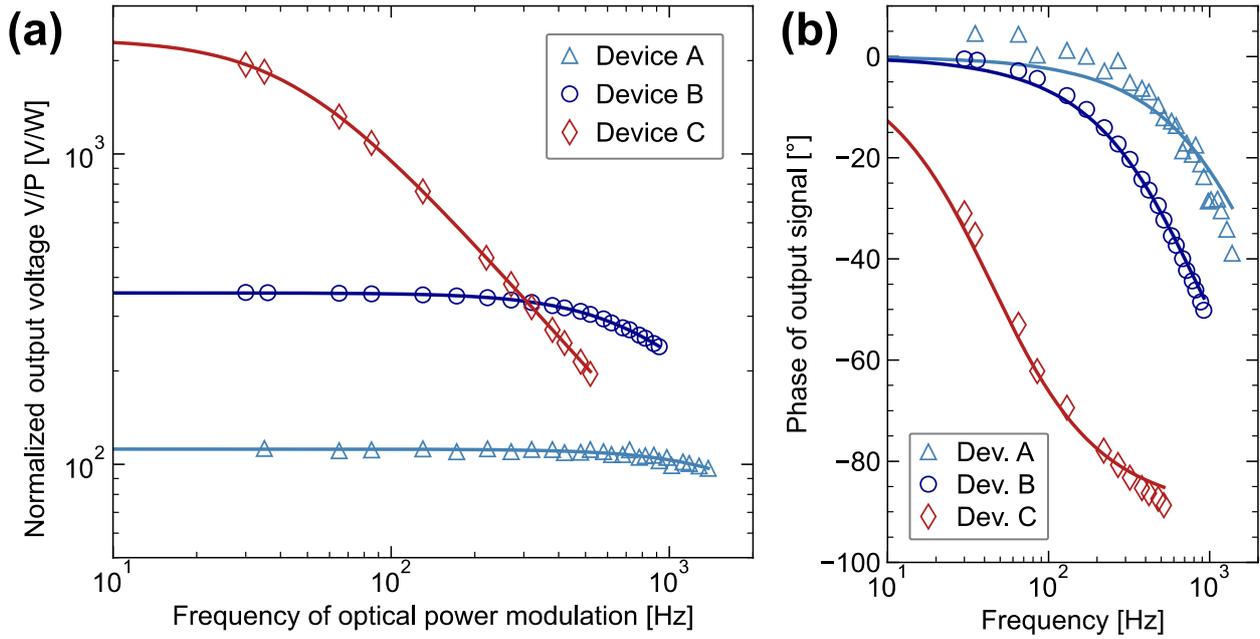

Figure 2. The measured magnitude (a) and phase (b) of the frequency response of the infrared detectors. In (a), the detector output voltage V is normalized by the optical power incident on the absorber area P. This normalization corresponds to the frequency-dependent responsivity. The response model of eqs. (1) and (2) was fitted to the measured detector output voltage amplitudes as a function of the frequency of the optical power modulation. The lines show the calculated model data. The fitted thermal time constants $\tau$ are listed in Table 1. The other fitting parameters for Devices A, B, and C are $V_{ampl}$ = 1.388 ± 0.006 µV, 3.740 ± 0.004 µV, and 63.72 ± 0.07 µV, respectively. Due to the higher optical power needed in the higher frequency range (see the experimental section), Device A was measured using the blackbody infrared source at 1200 °C, and 800 °C was used for the other devices. The amplitude of the total optical power incident on the absorber area is 12 ± 5 nW (2.1 mW/cm$^2$), 10 ± 1 nW (1.7 mW/cm$^2$), and 27 ± 2 nW (1.7 mW/cm$^2$) in the measurements of Devices A, B, and C, respectively.

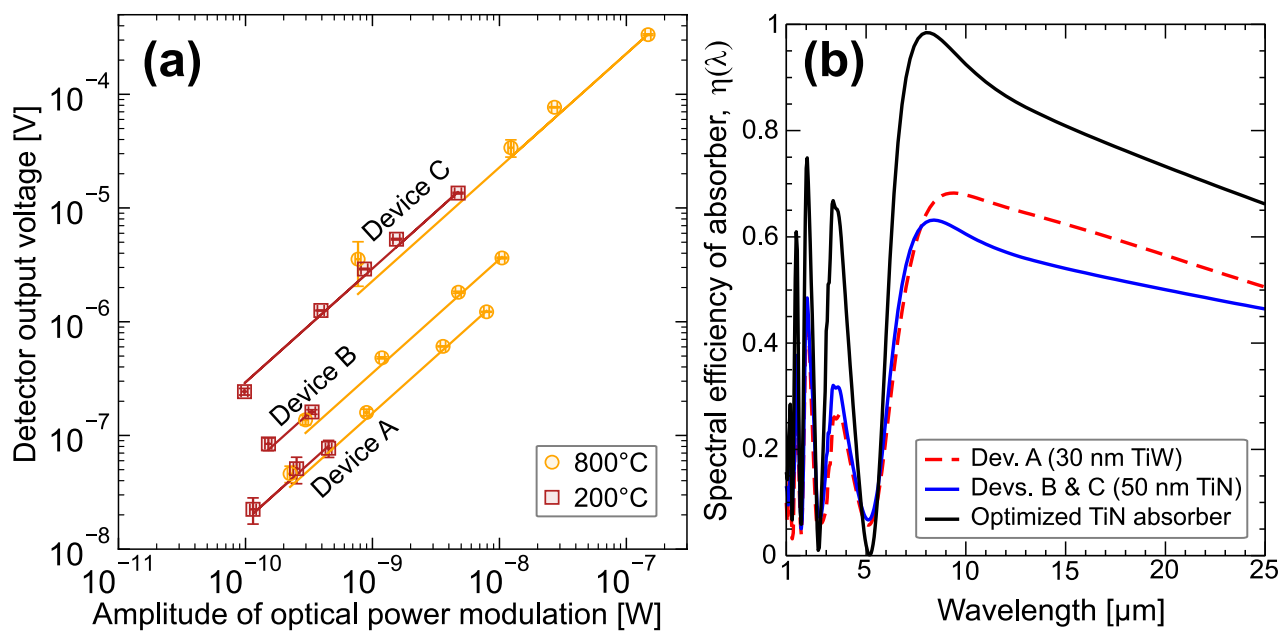

Figure 3. (a) The power responses of the detectors of Fig. 1 measured using various temperatures and output apertures of the blackbody infrared source. The power amplitude is the total optical power incident on the absorber area. The linear fits to the data (solid lines) correspond to the responsivities listed in Table 1. (b) Simulated spectral efficiencies of the absorbers of the detectors, $\eta_{abs}(\lambda)$, as a function of the optical signal wavelength (calculated for the unit cells of the absorber grids with periodic boundary conditions). The optimized TiN absorber (black solid line) utilizes the TiN thickness of 24 nm and a perfect mirror as the back reflector.

20 (20)

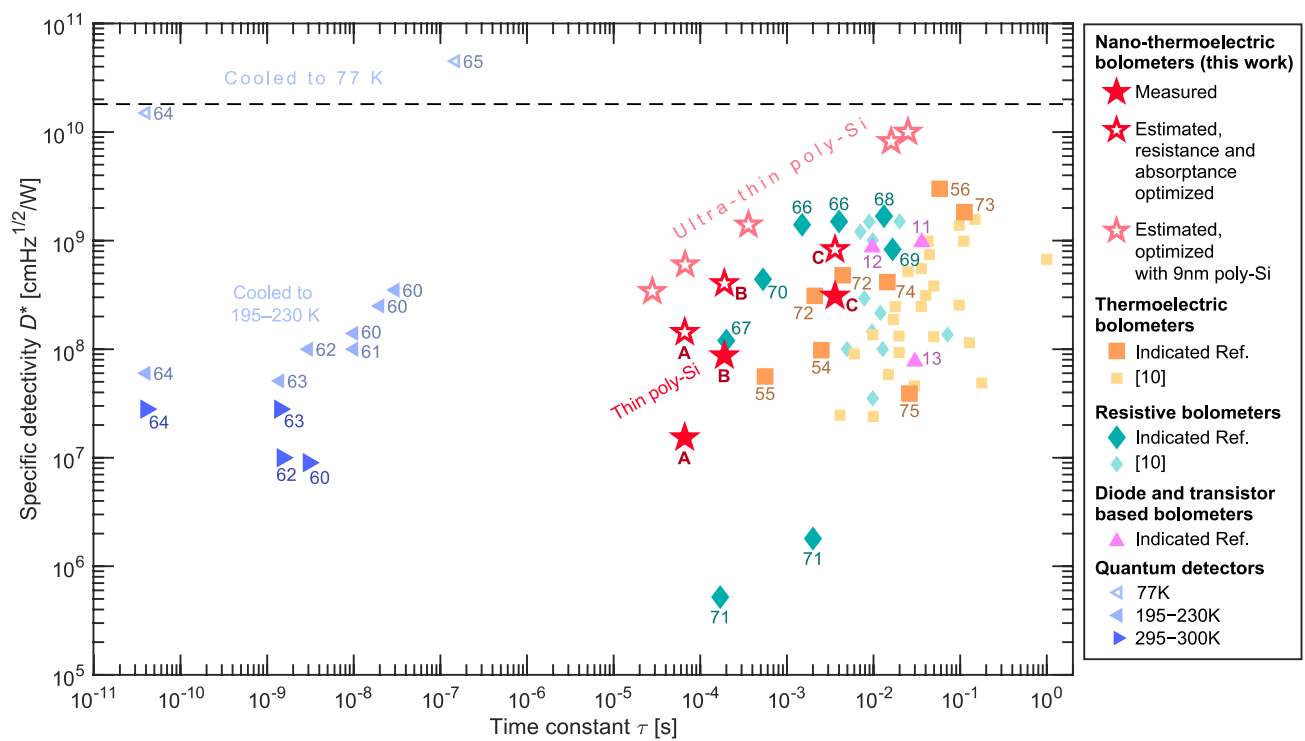

Figure 4. Time constants τ and specific detectivities D* of the nano-thermoelectric bolometers presented in this work and of commercial and non-commercial LWIR bolometers and quantum LWIR detectors [10–13,54–56,60–75]. The data of this work includes the measured and estimated detectors based on 70 – 80 nm poly-Si (the estimates calculated with beam resistance and unity absorptance), and the estimated ones with 9 nm poly-Si and different beam and absorber geometries (see Supplementary Material). The dashed black line shows the 300 K background radiation limit for thermal detectors calculated using full spectrum in the half space case [76]. The D* values of Refs. [13] and [75] were calculated using the reported responsivity, NEP, and geometrical data as described in Supplementary table 1, which includes the spectral ranges of the detectors of the publications as well.